\documentclass[fleqn,usenatbib]{mnras}

\usepackage{newtxtext,newtxmath}
\usepackage{xcolor}

\usepackage[T1]{fontenc}
\usepackage{ae,aecompl}

\usepackage{graphicx}	
\usepackage{amsmath}	
\usepackage{soul} 

\usepackage[utf8]{inputenc}
\usepackage[T1]{fontenc}
\usepackage[english]{babel}
\usepackage{soul}
\usepackage{mhchem}

\title[The Si+SO2 reaction]{
The Si+SO$_2$ collision and an extended network of neutral-neutral reactions between silicon and sulphur bearing species.}

\author[D. R. Campanha et al.]{
Danilo R. Campanha$^{1}$
Edgar Mendoza,$^{2}$ 
Mateus X. Silva,$^{1}$
Paulo F. G. Velloso,$^{1}$\newauthor
Miguel Carvajal,$^{2,3}$
Valentine Wakelam,$^{4}$
Breno R. L. Galvão$^{1}$  \thanks{Contact e-mail: \href{mailto:brenogalvao@gmail.com}{brenogalvao@gmail.com (BRLG)}}
\\
$^{1}$Departamento de Química, Centro Federal de Educação Tecnológica de Minas Gerais, CEFET-MG \\
Av. Amazonas 5253, 30421-169, Belo Horizonte, Minas Gerais, Brazil \\
$^{2}$Dept. Ciencias Integradas, Facultad de Ciencias Experimentales, Centro de Estudios Avanzados en Física,\\ Matemática y Computación,Unidad Asociada GIFMAN, CSIC-UHU, Universidad de Huelva, Spain\\
$^{3}$Instituto Universitario Carlos I de F\'{\i}sica Te\'orica y Computacional, Universidad de Granada, Spain\\
$^{4}$Laboratoire d’astrophysique de Bordeaux, Univ. Bordeaux, CNRS, B18N, allée Geoffroy Saint-Hilaire, 33615 Pessac, France. 
}

\date{Accepted XXX. Received YYY; in original form ZZZ}

\pubyear{2015}

\begin{document}
\label{firstpage}
\pagerange{\pageref{firstpage}--\pageref{lastpage}}
\maketitle

\begin{abstract}

The Si+SO$_2$ reaction is investigated to verify its impact on the abundances of molecules with astrochemical interest, such as SiS, SiO, SO and others. According to our results Si($^3$P) and \ce{SO2} react barrierlessly yielding only the monoxides SO and SiO as products. No favourable pathway has been found leading to other products, and this reaction should not contribute to SiS abundance.  Furthermore, it is predicted that SiS is stable in collisions with \ce{O2}, and that S($^3$P)+\ce{SiO2} and O($^3$P)+OSiS will also produce SO+SiO.
Using these results and gathering further experimental and computational data from the literature, we provide an extended network of neutral-neutral reactions involving Si- and S-bearing molecules. The effects of these reactions were examined in a protostellar shock model, using the {\sc Nautilus} gas-grain code.
This consisted in simulating the physicochemical  conditions of a shocked gas evolving from {\it i.} primeval cold core, {\it ii.} the shock region itself, {\it iii.} and finally the gas bulk conditions after the passage of the shock. Emphasising on the cloud ages and including systematically these chemical reactions, we found that [SiS/H$_2$] can be of the order of $\sim$ 10$^{-8}$ in shocks that evolves from clouds of $t=1\times 10^6$~yr, whose values are mostly affected by the SiS+O $\longrightarrow$SiO+S reaction. Perspectives on further models along with observations are discussed in the context of sources harbouring molecular outflows. 

\end{abstract}


\section{Introduction}

Although most of interstellar silicon is stored in the core of dust grains, at their surface atomic Si may be hydrogenated and converted to silane (SiH$_4$). Through this mechanism, a small portion of this element can be released in the gas phase, through desorption or sputtering~\citep{mackay1995chemistry}, and later converted into simpler species by photodissociation in the gas phase. 
In fact, silicon-bearing molecules are an important class of species in late type stars and star forming regions~\citep{MCC03:697}. For example, SiO masers have been used to investigate the structure and dynamics of such regions.

Contrary to the fairly well understood SiO chemistry, SiS formation and reactivity is still an open research topic, and new neutral-neutral reactions are being investigated recently, both theoretically~\citep{ROS18:695,PAI20:299,MOT21:37} and experimentally~\citep{DOD21:7003}. 
Bridging these two major silicon bearing species, the OSiS molecule has also received attention, and~\cite{ ESP13:A123} have provided an upper limit for its column density in Orion KL.
Its gas  phase spectrum has been obtained experimentally~\citep{Tho11:1228}, and the search for this molecule in the interstellar medium is an ongoing topic~\citep{PAI18:1858}.

A computational investigation~\citep{ZAN18:38} has predicted that the rate coefficient for the Si+SO\ce{-> SiO +S} reaction is one order of magnitude higher than Si+SO\ce{-> SiS +O}, and also that SiS is efficiently destroyed by collisions with atomic oxygen. Assuming that  Si+\ce{SO2} would produce SiO and SiS with the same rate coefficient, this same study proposes that Si+\ce{SO} and Si+\ce{SO2} are main sources of SiS, even though the latter reaction has not been investigated computationally or experimentally. Indeed, \ce{SO2} is one of the major sulphur bearing molecules~\citep{UMIST2006} in the insterstellar medium (ISM), and it would be important to verify if the Si+\ce{SO2} reaction is in fact a source of SiO, SO, and SiS. The only neutral-neutral destruction routes of \ce{SO2} that have been added to major astrochemistry databases such as KIDA~\citep{KIDA} and UMIST~\citep{UMIST} are collisions with atomic C, O, S, and H.

Regarding the SiS compound, \citet{Morris1975} reported its first detection in the molecular envelope of IRC+10216, a well studied carbon star. They observed the SiS $J$=5--4 and $J$=6--5 transitions at $\sim$~90771.85 and 108924.64~MHz, respectively. In addition to IRC+10216, they also searched for SiS in other 11 sources, however the molecule was not  detected. \citet{Ziurys1988} carried out observations towards the Orion Kleinmann-Low (KL) nebula reporting the detection of the SiS $J$=5--4 and $J$=6--5  transitions, whose spectral line profiles were associated with a moderate velocity outflow. Thus, the SiS production was discussed as being \lq\lq shock enhanced\rq\rq \ and favoured by the high and broad distribution of temperatures and densities present in the outflow. Using the same SiS $J$=5--4 and $J$=6--5 transitions,  \citet{Ziurys1991} reported new detections of SiS in three regions known for harbouring warm outflows sources: W51, Orion-S and Sgr-B2(N). In that work, it was discussed the relationship among SiO, SiS and the needed thermal conditions to explain how those species might be present in sources as outflows and dark clouds. In a more recent study, \citet{podio2017silicon} used the IRAM-30m and PdBI observatories to analyse Si-bearing molecules in the protostellar shock L1157-B1. They detected SiS by means of six rotational lines, including the $J$=5--4 and $J$=6--5 transitions detected in the previous works. Thus, the rotational diagram of SiS yielded temperatures and column densities of $\sim$~24~K and 4.3 $\times $10$^{13}$~cm$^{-2}$, respectively. They also found that SiO and SiS have a different spatial distribution and, therefore, different chemical origins. 

In the context of evolved stars,  \citet{Gong2017} reported the presence of a SiS (1--0) maser in the inner region of IRC+10216. They also observed quasi-thermal emission of the same molecule but in the extended circumstellar envelope of the evolved star. \citet{Velilla2017} observed rotational lines in vibrationally excited states of SiS and SiO isotopologues towards the molecular envelope of the oxygen rich AGB star IK Tau. They found rotational temperatures between 15 and 40~K for various molecules. However, NaCl and SiS displayed a higher temperature of $\sim$~65~K. From SO$_2$ emission, they also found a warm component with $T_{rot} \sim$~290~K. 
In a recent work, \citet{Rizzo2021} carried out a search for SiO emission in a sample of 67 evolved oxygen rich stars; apart from the detection of molecules like SiO, SO and H$_2$S, they detected transitions of SiS in various  sources.

This work is focused in investigating the Si+\ce{SO2} reaction by computational tools, to assess its major chemical products. Using the similar tetratomic complexes, we also provide insights into the destruction of SiS by molecular oxygen and the reactions S+\ce{SiO2} and O+OSiS. As our final goal is to better describe the chemistry of silicon and sulphur bearing molecules in space, we use the present results, as well as computational and experimental data from the literature, to provide an extended network of neutral-neutral chemical reactions that may play a significant role to the evolution of these molecules. Such new network is then used in an astrochemical model of a shocked region, taking into account the physical conditions of L1157-B1 since the most recent detection of SiS was confirmed in that source. Thus, the impact of the new reactions is discussed in the context of that source. Perspectives on how to improve such models are also discussed.

\section{Computational Methods}
\label{sec:method}

The reaction mechanisms have been explored with DFT calculations using the M06-2X functional \citep{ZHA08:215} with the pcseg-2 basis set~\citep{jen14:1074}  and employing restricted open shell orbitals. Geometries optimisations, zero-point energies (ZPEs), vibrational frequencies and intrinsic reaction coordinate calculations were perfomed using the GAMESS-US package~\citep{GAMESS},  with a convergence criterion of $1\times 10^{-4}E_h a_0^{-1}$ for the geometry optimizations and $1\times 10^{-5}E_h$ for the self consistent-field iterations. 
All structures were double checked with the unrestricted calculations with the ~$\omega$B97X-D~\citep{CHA08:6615} functional to avoid reporting spurious structures.  
The two functionals agree well,  with an average deviation of $8\rm kJ\,mol^{-1}$, and a comparison between the two is given in the Supporting Information.

At the M06-2X optimised geometries, single-point energy calculations were performed using the explicitly correlated coupled cluster method (CCSD(T)-F12)
method~\citep{ADL07:221106,KNI09:054104}, with the aug-cc-pV(T+d)Z basis set~\citep{DUN01:9244}  and employing the MOLPRO package \citep{MOLPRO}, which greatly enhances the accuracy of the energies reported and generally provides results within  $5\rm kJ\,mol^{-1}$ even for barrier heights~\cite{ZHA12:3157}. The final energies reported in the manuscript will therefore be termed CCSD(T)-F12/aug-cc-pV(T+d)Z//M06-2X/pcseg-2+ZPE(M06-2X/pcseg-2).

\section{Computational  Results}
\label{sec:results}

\subsection{Reaction products}

The collision of atomic silicon in its ground state ($^3$P) with \ce{SO2} can occur only in the triplet state of the tetratomic system, and may give rise to four possible exothermic:

\begin{align}
\label{eq:p1} \rm Si(^3P) + SO_2(^1A_1) \rightarrow & \rm SiO(^1\Sigma^+) + SO(^3\Sigma^-)\,\,\,\, \rm -247 \,kJ\,mol^{-1}\\
\label{eq:p2} \rightarrow & \rm  S(^3P)  + SiO_2(^1\Sigma_g)\,\,\,\,\,\,\,\, \rm -164 \,kJ\,mol^{-1} \\
\label{eq:p3} \rightarrow & \rm  SiS(^1\Sigma^+)  + O_2(^3\Sigma_g^-)\,\,\,\,\, \rm -31 \,kJ\,mol^{-1}\\
\label{eq:p4} \rightarrow & \rm  O(^3P)  + OSiS(^1\Sigma^+)\,\,\,\,\,\, \rm -23 \,kJ\,mol^{-1}
  \end{align}

Note that all products are in their respective ground state, and no intersystem crossings with a possible singlet potential energy surface is necessary to this study, as they would lead to higher energy excited states of the products. 
Even though this work provides (section 5) a simple model of a protostellar shock, the computational results presented in this section are applicable to diverse astronomical environments, and the relative importance of each reaction will depend on the abundance of the reactants in the gas-phase in a given region. For these reasons, possibilities other than shocks are described in the following paragraphs.

As can be seen, the most exothermic products that can be obtained is the formation of silicon and sulphur monoxides, both of which have been detected in the interstellar medium, and this reaction could be an important source for them in regions with a significant \ce{SO2} abundance.  Reaction~(\ref{eq:p2}) provides a source of \ce{SiO2}, whose gas phase chemistry (together with SiO) has been proposed by~\cite{YAN18:774} to drive an exothermic chemistry yielding larger silicon oxides, which will ultimately be converted to silicate grains.  This is a very important topic, for we still do not have a detailed chemical description of the formation of such grains.

Reaction~(\ref{eq:p3}) has been proposed to be one of the main sources of SiS in a post-shocked gas~\citep{ZAN18:38}, and since there is no theoretical calculation or experimental evidence for this, one of the goals of this work is to provide a computational verification for this route. Finally, reaction~\ref{eq:p4} will lead to the exotic OSiS molecule. Although this species has not been detected in the ISM,  \cite{ESP13:A123} have estimated an upper limit for its abundance in Orion KL. Its experimental rotational spectrum is already measured~\citep{Tho11:1228}, and theoretical predictions have also been performed~\citep{PAI18:1858}.

Even though the energetic analysis of this section indicates that SiO and SO may be the most important product of the title reaction, the presence of reaction barriers, or indirect mechanisms involving multiple isomerisations may give rise to unexpected branching ratios. For this reason, the whole PES is  analysed in the next section.

\subsection{SO and SiO formation}

\begin{figure}
\begin{center}
\includegraphics[angle=0,width=0.47\textwidth]{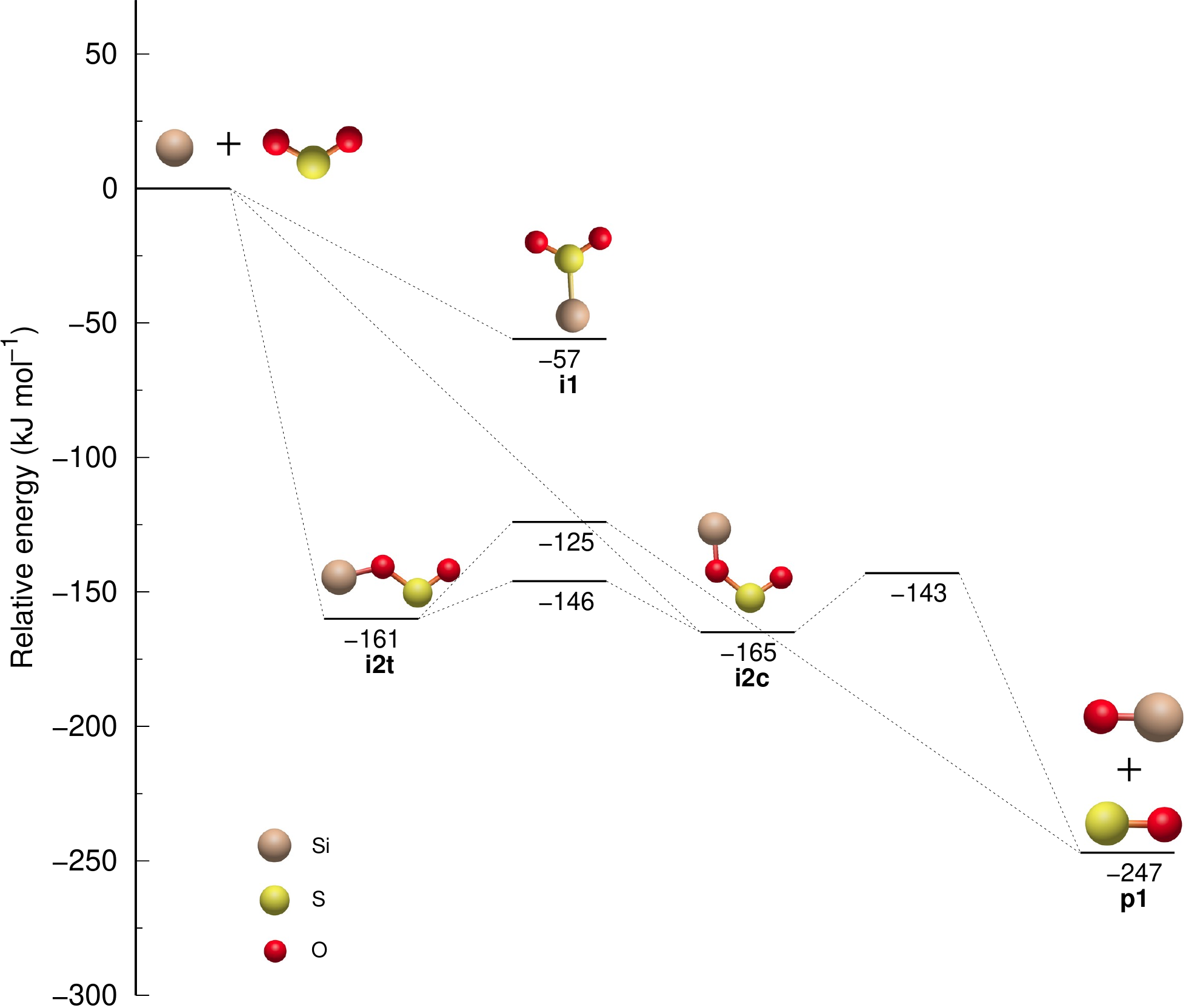}
\caption{\label{fig:PESp1} Potential energy profile for $\rm Si(^3P) + SO_2(^1A_1)$ collision at the CCSD(T)-F12/aug-cc-pV(T+d)Z//M06-2X/pcseg-2+ZPE(M06-2X/pcseg-2) level relative to reactants. Atoms are coloured as follows: oxygen (red); silicon (brown); sulfur (yellow).}
\end{center}
\end{figure}

The incoming Si atom may approach the SO$_2$ molecule via either the oxygen or sulphur ends. As can be seen in Fig. \ref{fig:PESp1}, both attacks occur barrierlessly and lead to different energy minima. The attack to the central atom leads to the {\bf i1} intermediate (\ce{SiSO2}), and no isomerisation or dissociation path has been found from it, and thus a trajectory leading to this  minimum will have only the recrossing option (back to reactants).
The most energetically favoured attack is to the oxygen of the \ce{SO2} molecule, from where both cis and trans SiOSO intermediates are achieved with similar energies (hereafter denoted {\bf i2}t and {\bf i2}c). The isomerisation between the two species can occur easily, through a rather small energy barrier. Both structures can dissociate forming sulphur and silicon monoxides (reaction ~\ref{eq:p1}) passing through   transition states that lie below the reactants limit and, thus, the reaction ~\ref{eq:p1} should occur easily at any temperature.
Relative to the energy of each intermediate, the dissociation barriers from the trans and cis isomers are  36 and 22 kJ mol$^{-1}$, respectively. From the cis isomer, we have also unravelled a transition state for a Si atom migration from one oxygen to another, connecting two symmetrically equivalent {\bf i2}c minima. Since this is not important from the astrochemical point of view, this is only included as supporting information.

Despite our efforts, we have found no mechanisms leading  the Si+SO$_2$ reaction to any other possible products. This indicates that the reaction~\ref{eq:p1} will be the only product of the title collision. Without sharing the trajectories with any other products, the rate coefficient for  this reaction will likely approach the gas kinetic limit of a few $10^{-10}\rm cm^3 s^{-1}$. 
In the next  subsections, we explore the outcome of the other  possible collisions: \ce{S + SiO2}, \ce{SiS + O2} and \ce{O + OSiS}.

\subsection{S+SiO$_2$ and O+OSiS collisions}

\begin{figure}
\begin{center}
\includegraphics[angle=0,width=0.47\textwidth]{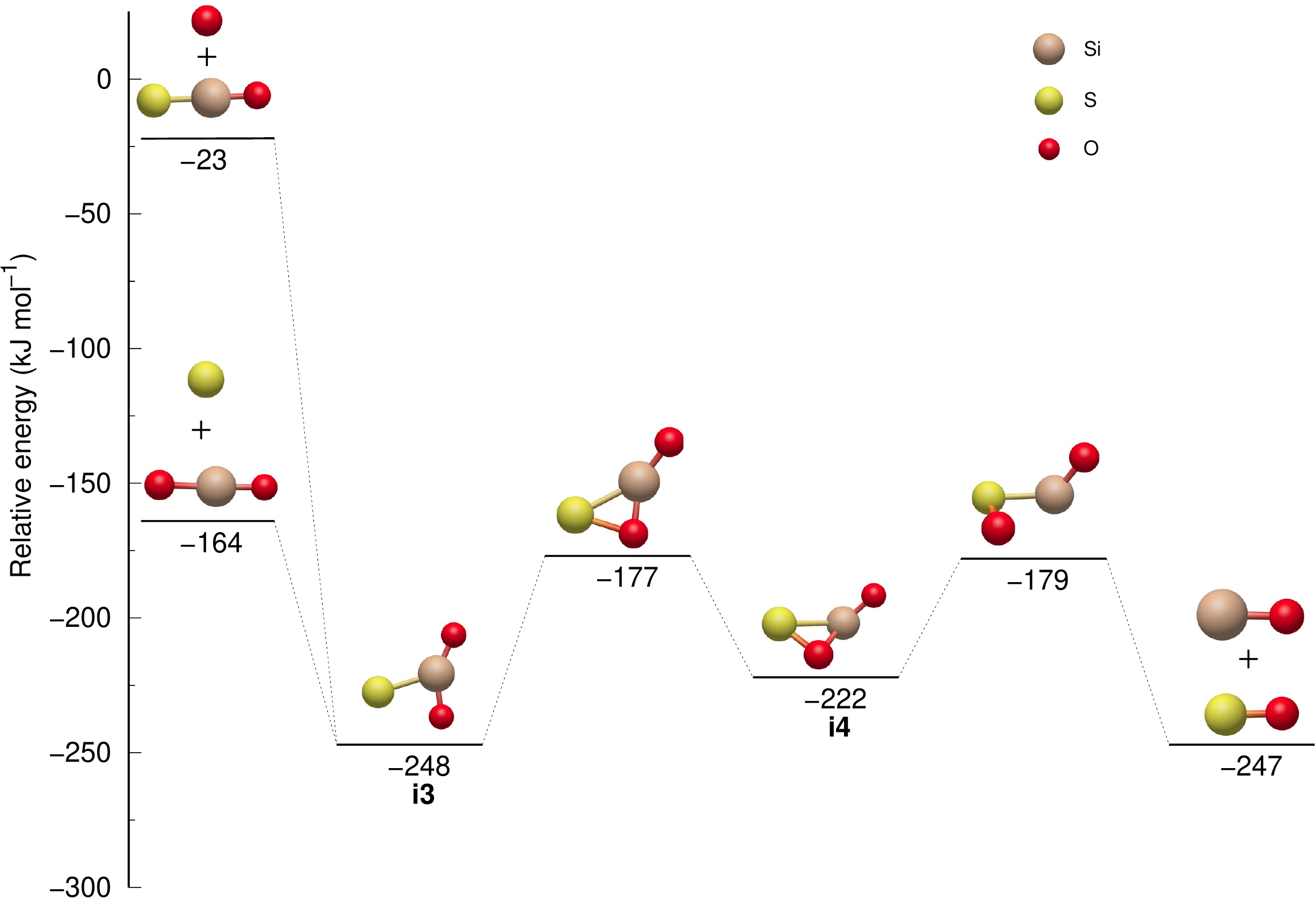}
\caption{\label{fig:PESp2} Potential energy profile for $\rm S(^3P)  + SiO_2(^1\Sigma_g)$ and $\rm O(^3P)  + OSiS(^1\Sigma^+)$ collision. Energies are given relative to  \ce{Si + SO2}. Atoms are coloured as in Fig.~\ref{fig:PESp1}.}
\end{center}
\end{figure}

Although reactions (\ref{eq:p2}) and (\ref{eq:p4}) are exothermic, these products can not be achieved from the intermediates {\bf i1} or {\bf i2} that follow naturally from the \ce{Si + SO2} interaction, as was seen in Fig.~\ref{fig:PESp1}. In order to show how \ce{S + SiO2} and \ce{O + OSiS} are related to this PES, Fig.\ref{fig:PESp2} provides their connections to different structures. First of all, both interactions occur barrierlessly and achieve the same intermediate {\bf i3}, a planar form of \ce{SSiO2}. This can then isomerize to the non planar {\bf i4} isomer, from where dissociation to the monoxides SiO and SO can occur again.

This is a quite relevant result by itself, as it shows that both \ce{S + SiO2} and \ce{O + OSiS} reactions will quickly lead to the monoxides \ce{SiO + SO} under astrochemical conditions, even at low temperature regions. This could be an explanation for the non detection of OSiS so far, as this species will be depleted by oxygen atoms, and the search for it should be focused on regions with lower oxygen abundances. Furthermore, as \ce{SiO2} may be a building block to silicate grains, it is relevant to point out that the product \ce{S + SiO2} is an efficient sink for this important molecule.

Note that the barrier from {\bf i3} to {\bf i4}, a crucial step in the mechanism described above, lies only 13 $\rm kJ\, mol^{-1}$ below the \ce{S + SiO2}. The pure DFT calculations provide lower values for this barrier, lying 18 and 16 $\rm kJ\, mol^{-1}$ below \ce{S + SiO2} for the M06-2X and $\omega$B97X-D functionals, respectively. This is important, for if this barrier lain above \ce{S + SiO2} it would mean that this reaction would not be opened for low temperatures.

\subsection{Destruction of SiS by molecular oxygen}

We have previously shown that reaction~(\ref{eq:p3}) cannot occur, even though it is thermodynamically favourable.
However, it would be instrumental in knowing if molecular oxygen can destruct, as occurs with atomic oxygen~\citep{ZAN18:38}. In fact, as it will be shown later, the \ce{SiS + O} reaction has a major importance on the evolution of SiS in space.

\begin{figure}
\begin{center}
\includegraphics[angle=0,width=0.45\textwidth]{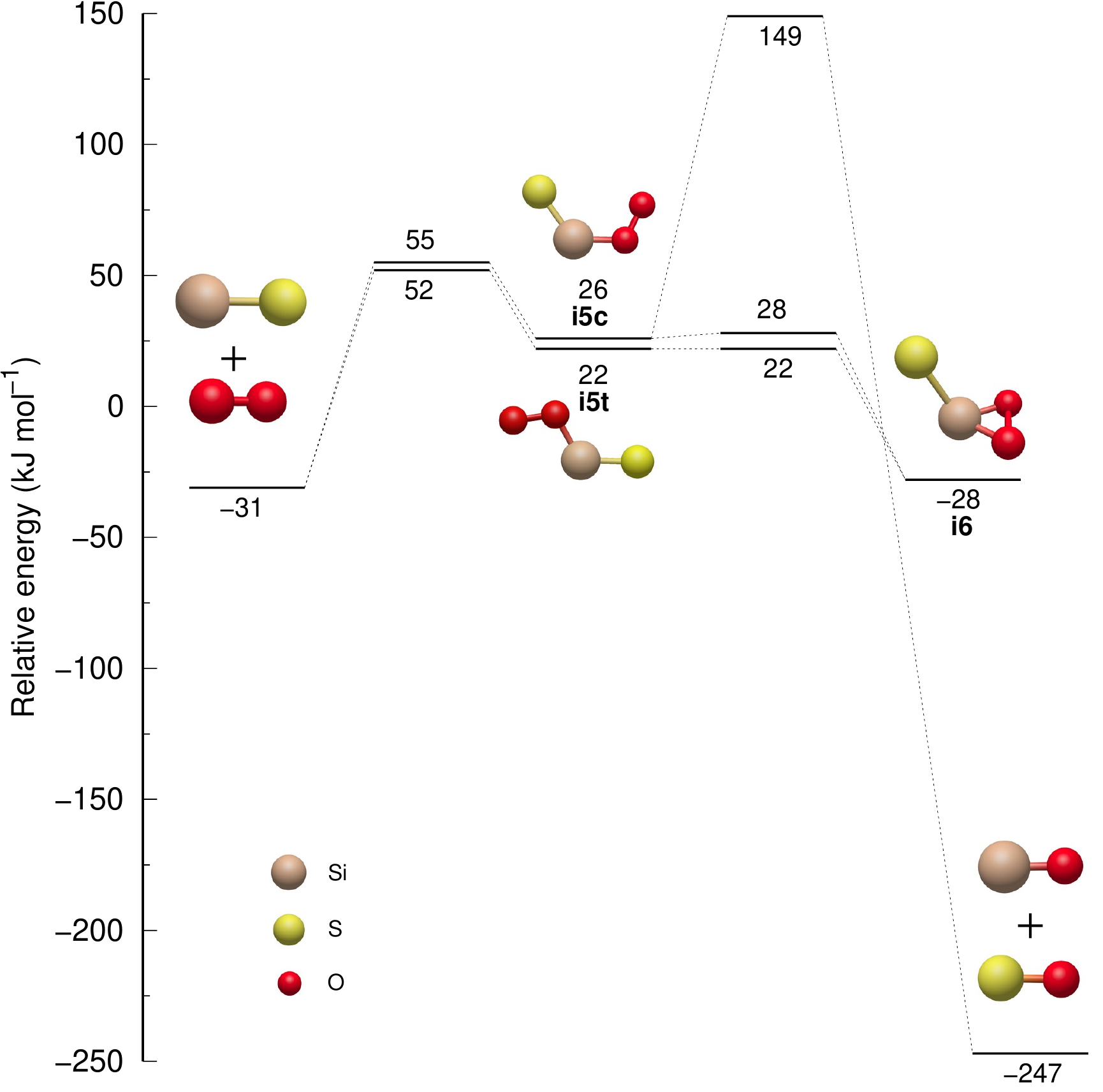}
\caption{\label{fig:PESp3} Potential energy profile for $\rm SiS(^1\Sigma^+)  + O_2(^3\Sigma_g^-)$ collision. Energies are given relative to  \ce{Si + SO2}. Atoms are coloured as in Fig.~\ref{fig:PESp1}.}
\end{center}
\end{figure}

We have checked that the initial interaction between $\rm SiS(^1\Sigma^+)  + O_2(^3\Sigma_g^-)$ is always repulsive, and for this reason this SiS oxidation mechanism is unlikely to play a role in the destruction of this species. In fact, from our calculations (Fig.~\ref{fig:PESp3}), when the oxygen molecule attacks SiS via its silicon end, two potential energy minima can be obtained: one trans OOSiS ({\bf i5}t) and one cis ({\bf i5}c). However, the barriers at 52 and 55 kJ mol$^{-1}$ (83 and 86 $\rm kJ\,mol^{-1}$, with respect to SiS+\ce{O2}) make this formation unlikely in cold regions. These minima can either isomerize to the non planar {\bf i6} structure, or dissociate towards SiO+SO, but this latter reaction presents an even higher activation energy. The overall barrier for \ce{SiS + O2 -> SiO + SO} is 180~$\rm kJ\,mol^{-1}$.

\begin{figure*}
\begin{center}
\includegraphics[angle=0,width=0.75\textwidth]{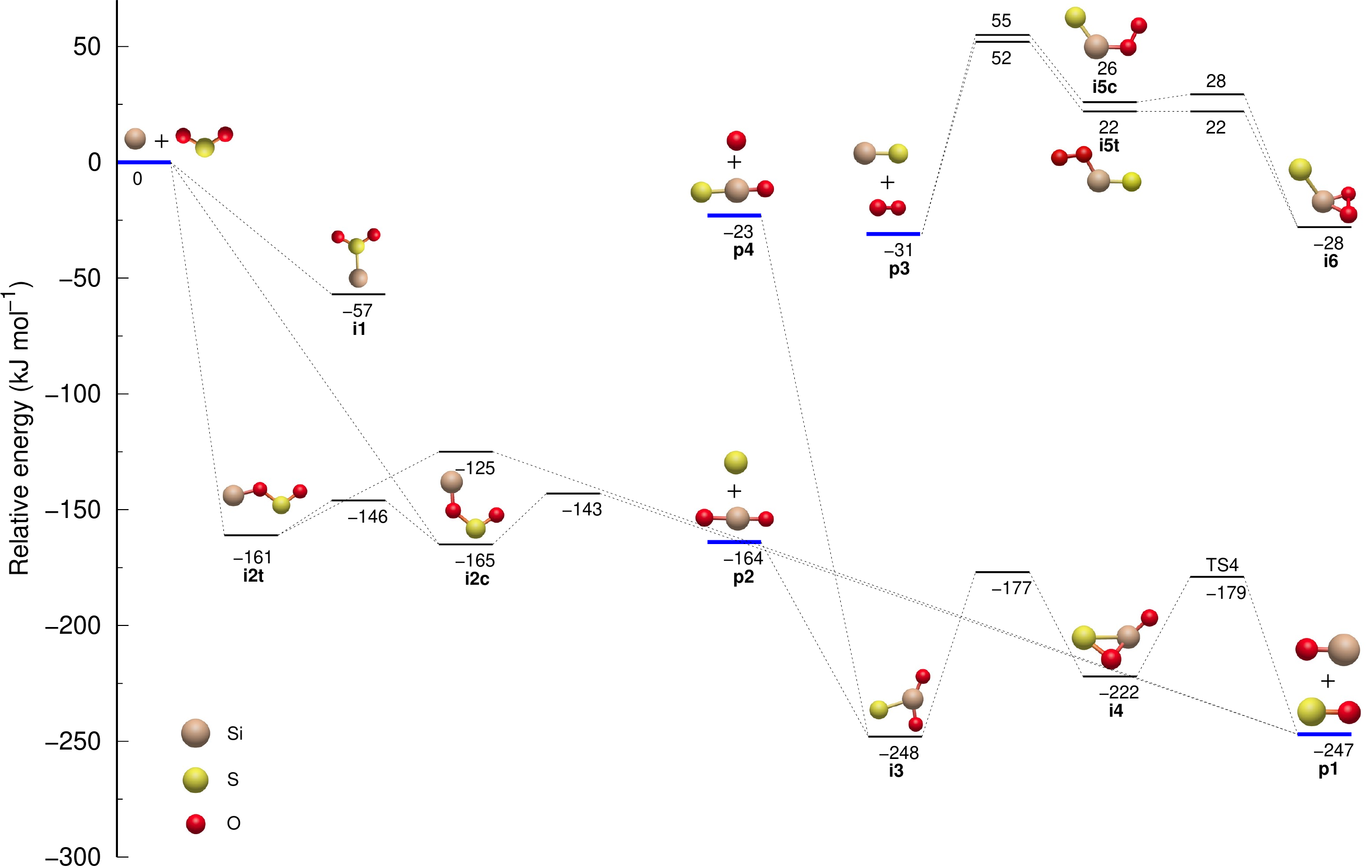}
\caption{\label{fig:PESall} Potential energy profile for the $^3\rm SiSO_2$ tetratomic system as calculated at the CCSD(T)-F12/aug-cc-pV(T+d)Z//M06-2X/pcseg-2+ZPE(M06-2X/pcseg-2). Energies are given relative to  \ce{Si + SO2}. The asymptotic channels are highlighted in blue.}
\end{center}
\end{figure*}

Finally a complete overview of the whole $^3\rm SiSO_2$ tetratomic system is given in Fig.~\ref{fig:PESall}, where all intermediates and mechanisms explored in this work can be seen together. The {\bf i5}c \ce{-> SiO + SO} barrier has been omitted due to its high energy. For clarity, the connection between \ce{SiS + O2 -> SiO + SO} is  not evinced here because it is already shown in Fig.~\ref{fig:PESp3}.

\section{Extended Chemical Network}

Our ultimate goal is to shed light on the chemistry of silicon and sulphur bearing molecules in the interstellar medium. 
For this, all relevant reactions should be included in astrochemical models, which currently lack many important neutral-neutral processes. 
Having established new relevant data in the previous sections, we now gather data from several reactions that have been explored experimentally or computationally available in the literature on neutral-neutral reactions for inclusion in astrochemical databases. The rate coefficients of such reactions are given in Table~\ref{tab:net}.

\begin{table*}
	\caption{Extended network for silicon and sulphur neutral-neutral reactions.}
	\label{tab:net}
    \centering
	\begin{tabular}{lcccc}
		\hline
		\textbf{Reaction} & $\alpha \rm (cm^{-3}s^{-1})$ & $\beta$ & $\gamma \rm (K)$ & description$^a$\\
		\hline
\textbf{SiS formation}\\		
\ce{Si +SH}$\rightarrow$ \ce{SiS + H} & $0.77\times 10^{-10}$ & -0.756 & 9.873 & QCT,~\citep{MOT21:37} \\

\ce{Si +H2S}$\rightarrow$ \ce{SiS + H2} & $1\times 10^{-10}$ & 0 & 0 & Exp.,~\citep{DOD21:7003} \\		

\ce{Si +SO}$\rightarrow$ \ce{SiS + O} & $1.77\times 10^{-11}$ & 0.16 & -20 & QCT,~\citep{ZAN18:38} \\		

\ce{SiH +S}$\rightarrow$ \ce{SiS + H} & $1\times 10^{-10}$ & 0 & 0 & PES.~\citep{ROS18:695} \\	

\ce{SiH +S2}$\rightarrow$ \ce{SiS + SH} & $1\times 10^{-10}$ & 0 & 0 & PES.~\citep{ROS18:695} \\

\textbf{SiS destruction}\\
\ce{SiS +O}$\rightarrow$ \ce{SiO + S} & $9.53\times 10^{-11}$ & 0.29 & -32 & QCT~\citep{ZAN18:38} \\	

\textbf{Reactions between Si- and S-bearing species}\\	
\ce{Si +SO}$\rightarrow$ \ce{SiO + S} & $1.53\times 10^{-10}$ & -0.11 & 32 & QCT~\citep{ZAN18:38} \\	
\ce{Si +SO2}$\rightarrow$ \ce{SiO + SO} & $1\times 10^{-10}$ & 0 & 0 & PES, this work \\	
\textbf{Reactions not destroying SiS}\\
\ce{SiS +H} & 0 & 0 & 0 & Endo.,~\citep{ROS18:695} \\	
\ce{SiS +H2} & 0 & 0 & 0 & Endo.,~\citep{PAI20:299} \\	
\ce{SiS +O2} & 0 & 0 & 0 & this work, see discussion \\	
		\hline
	\end{tabular}
	
{$^a$ QCT: calculated by quasi-classical trajectories simulations; Exp: inferred from crossed molecular beams experiment; PES: inferred from barrierless in the potential energy surface  
; Endo: Calculated to be endothermic by more than 250 kJ\, mol$^{-1}$.}
\end{table*}

\section{Astrochemical modelling}

Quantum calculations and astrochemical simulations have demonstrated to be a powerful tool to study the gas-grain processes that lead  to the synthesis of molecules in the ISM. Along with radiative transfer models, those methods contribute to predict, e.g., molecular column densities and synthetic spectra in the sub-mm domain (e.g. \citealt{Puzzarini2020,Jinjin2021,Gavino2021,Keil2022}). In this work, we address specific questions on the chemistry of Si- and S-bearing molecules in the context of protostellar shock regions, taking L1157-B1 as a template source since molecules as SiS, SiO, HS and H$_2$S were already detected towards it \citep{podio2017silicon,Holdship2019}.  Thus, we adopt a theoretical approach to be used in follow-up studies on the observation of Si- and S-containing molecules.

To do so, we use the {\sc Nautilus} gas-grain code which allows to compute rate equations, under different physical and chemical conditions, in order to predict molecular abundances in interstellar objects. The gas-phase chemistry is based on the kida.uva.2014 catalogue of chemical reactions \citep{KIDA}; in addition, physicochemical  processes of adsorption, desorption and chemistry of surfaces are also simulated by the code \citep[e.g.,][]{Ruaud2016}. In that sense, our simulation consists of a  state-of-the-art 3-phase model which includes 1) gas-phase, 2) surface and 3) bulk chemical species, where the latter ones depend on the grain layer in which they might be present. For instance, while diffusion and chemical reactions can occur both in the surface and bulk of grains, accretion and desorption processes are expected to happen only on the surface of grains \citep{Iqbal2018}.

In a previous work, \citet{MOT21:37} investigated the impact of the  reaction Si + HS $\longrightarrow$ SiS + H on the final abundance of SiS in the context of L1157-B1. The simulation yielded positive results since the computed abundances, of the order of [SiS/H$_2$]$\thickapprox$10$^{-8}$, were comparable to the observed value \citep{podio2017silicon}. Those authors observed this molecule at a rotational temperature $T_{rot} \thickapprox$ 24~K for a cavity region with kinetic temperatures between 25--80~K. \citet{podio2017silicon} also observed that SiS has a different gas distribution in comparison with SiO, suggesting that they are not chemically associated.

In order to address new questions about Si and S-bearing molecules, in this work we have extended the network of chemical reactions, taking into account the equations listed in Table~\ref{tab:net}. Therefore, in agreement with \citet{podio2017silicon}, our focus also consisted in testing chemical reactions which does not involve SiO as a potential precursor of SiS. We have used the same protostellar shock model but focusing  on the chemical evolution of the primeval cloud. As a whole, the simulation consists in a three steps model, namely: $1^{st}$, an initial and long molecular cloud step; $2^{nd}$, the passage of the shock, a fast and energetic step; and $3^{rd}$, an evolved and shocked gas after the passage of the shock.  In the literature, detailed simulations have presented results about the gas-grain chemistry and sputtering of molecules in shocks (e.g., \citealt{Holdship2017,James2020,Lefloch2021}). In this work, it is worth mentioning that our goal is simply to compute abundances in warm environments and hence our modelling does not include any treatment of the source of the heating e.g. of the shock. Further aspects about how the gas velocity distribution and shock waves affect the SiS chemistry are beyond the scope of this work, although we expect to face it in  follow-up investigations. Taking into account that an interstellar cloud can evolve via different and complex ways, with direct implications on the chemistry and molecular abundances, we tested various ages during the first step of the model, including time scales from 1 $\times$ 10$^5$ to 1 $\times$ 10$^6$~yr.

\subsection{Lifetime of the primeval molecular cloud}
\label{Sec:5.1}

The lifetime of clouds is a subject largely discussed in the literature that, along with molecular studies, provide key insights on the evolutionary phases of the ISM. Investigations suggest that cloud  lifetimes can go from few million years up to long ages of $\simeq$~10~Myr \citep{Koda2009}. The cloud lifetime not only depends on the gravitational collapse, it is also sensitive to the galactic dynamics, to processes such as the galactic shear, spiral arm interactions and cloud-cloud collisions, among others \citep{Jeffreson2018}. Recent ALMA observations of CO (1--0) in clouds, combined with theoretical approaches, suggested short timescales of the order of 1~Myr \citep{Kruijssen2019}. Nonetheless, in contrast to the argument that molecular clouds are short-lived and that star formation is rapid, investigations also suggest that cloud lifetimes can be as longer as $\simeq$ 10~Myr  \citep{Mouschovias2006}. From a chemical point view, studies on the detection and gas-grain formation of molecules are key aspects to understand the evolution of clouds. In a study based on the CO freeze-out onto grains, degree of molecular deuteration and the ortho-H$_2$ abundance in clouds, \citet{Pagani2011} found that the presence of DCO$^+$ in clouds would be in agreement with  a timescale of $\sim$~6$\times$10$^6$~yr. In interstellar cloud cores, \citet{Chang2007} performed gas-grain chemical simulations to estimate the mantle abundances of O-bearing molecules, they found  comparable values, at timescales of 2 $\times$ 10$^5$~yr, to those derived from observational methods in W33A and Elias 16. In an investigation based on a sample of clouds harbouring prestellar cores, from which a new generation of stars will form, \citet{Enoch2008} estimated lifetime values of 2-5 $\times$ 10$^5$~yr. To evaluate the impact of different cloud ages in our model, we found appropriate to compute the cloud ($1^{st}$) step using lifetimes between 0.1--1 $\times$ 10$^6$~yr. Ages longer than those are not favourable, e.g., to reproduce the rich chemistry observed toward protostellar objects; large integration times can also yield a fall in the gas-abundance values due to large accretion of molecules in grains \citep{Ruaud2016}.

\subsection{Shock model: chemical abundances and physical parameters}

In the first step of the model, we considered typical cloud conditions, for which it was adopted a gas and grain temperature of $T$=10~K, visual extinction $A_v$=10~mag, total H density of $n_{\text H}$=2$\times$10$^4$~cm$^{-3}$ and the standard cosmic ray ionisation rate $\zeta$=1.3 $\times$ 10$^{-17}$~cm$^3$ s$^{-1}$. Concerning the integration time, we present three results for the cloud timescales a) 1 $\times$ 10$^5$, b)   5 $\times$ 10$^5$ and c)  1 $\times$ 10$^6$~yr. As commented above, such ages are in agreement with the dynamic and collapse of clouds, and with the gas-grain chemistry of Si and S-bearing molecules observed in clouds, such as SiO, SO, SO$_2$, CS, OCS and H$_2$S \citep{Herpin2009,Agundez2013,Bilalbegovic2015,Neufeld2015}. The initial abundances used here are those listed in Table~\ref{table1}, whose values are in agreement with those observed in $\zeta$ Oph  \citep{Jenkins2009}. Those values were also used in previous works \citep{MOT21:37} to predict the SiS abundances in shock regions.

The second step corresponds to the shock itself, so that we ran \lq\lq fast\rq\rq \ models, to simulate the passage of the shock, using a quick integration time of 10$^2$~yr. The initial abundances used here were those obtained after running the (1$^{st}$) cold core model. Concerning the physical conditions, the gas density was kept of the same order but the temperature was increased to 100~K, which is assumed as a \lq\lq referential\rq\rq \ value to simulate the thermal effect responsible for the sublimation of volatile molecules from grains (e.g. \citealt{Potapov2019,Kalvans2020}); such gas temperature is also in agreement with the  kinetic temperatures derived from the molecular gas of L1157-B1 \citep{Lefloch2012,James2020}. In particular, \citet{Jaehyeock2017} performed experiments to analyse the interaction of SO$_2$ gas with a crystalline ice surface at low temperatures ($\sim$ 90--120~K), finding that thermal reactions can occur efficiently with the posterior desorption of molecules.

The third step represents the condition after the passage of the shock, generating a dense and quiet bulk of gas rich in organic molecules. In order to obtain the final abundances we ran the calculations whose starting point was the chemical conditions obtained from the precedent step. In order to mimic the conditions of a template shock such as L1157-B1, whose dynamical age is around 10$^3$~yr, we assume a dense condition, using $n_{\text H}$=2 $\times$ 10$^7$~cm$^{-3}$, and a cold regime for the gas and grain, adopting $T$=30~K, which is in agreement with the low temperatures obtained from SiS, SO and SO$_2$ observed in L1157-B1 \citep{POD17:L16,Holdship2019}. Here, instead of simulating a proper shock region, we are analysing the chemical effects of increasing both the gas density and temperature mimicking the physical conditions of a  protostellar shock in which SiS was observed. 
Concerning the dense condition mentioned above, it is worth mentioning that we adopted molecular densities of up to 2$\times$10$^7$~cm$^{-3}$ based on the SiO observations of L1157-B1 analysed by \citet{Spezzano2020}. In large velocity gradient calculations, those authors used H$_2$ densities from 1$\times$10$^3$ to 2$\times$10$^7$ cm${^-3}$, so that we have taken the highest density as a benchmark for the shock step computed here. We found it pertinent since density values of such order of magnitude can affect the chemistry of the molecules studied here. In further works, we will explore more comprehensively how molecular densities affect gas-grain chemical processes of Si- and S-bearing species.

\begin{table}
	\caption{Initial abundances used in the chemical model \citep{MOT21:37}.}
	\label{table1}
    \centering
	\begin{tabular}{lclc}
		\hline
		\textbf{Species} & \textbf{Abundance} & \textbf{Species} & \textbf{Abundance} \\
		\hline
		H$_2$  & 0.5                        &  Fe$^+$ & 2.0  $\times$ 10$^{-7}$ \\
		He     & 9.0  $\times$ 10$^{-2}$    &  Na$^+$ & 2.3 $\times$ 10$^{-7}$ \\
		N      & 6.2 $\times$ 10$^{-5}$     &  Mg$^+$ & 2.3 $\times$ 10$^{-6}$ \\
		O      & 3.3 $\times$ 10$^{-4}$     &  P$^+$  & 7.8 $\times$ 10$^{-8}$ \\
		C$^+$  & 1.8 $\times$ 10$^{-4}$     &  Cl$^+$ & 3.4 $\times$ 10$^{-8}$ \\
		S$^+$  & 1.5 $\times$ 10$^{-5}$     &  F      & 1.8 $\times$ 10$^{-8}$ \\
		Si$^+$ & 1.8 $\times$ 10$^{-6}$     &   \\
		\hline
	\end{tabular}
\end{table}

\section{Astrochemical model results}
\label{sec:6}

\begin{figure}
\begin{center}
\includegraphics[width=0.47\textwidth,keepaspectratio]{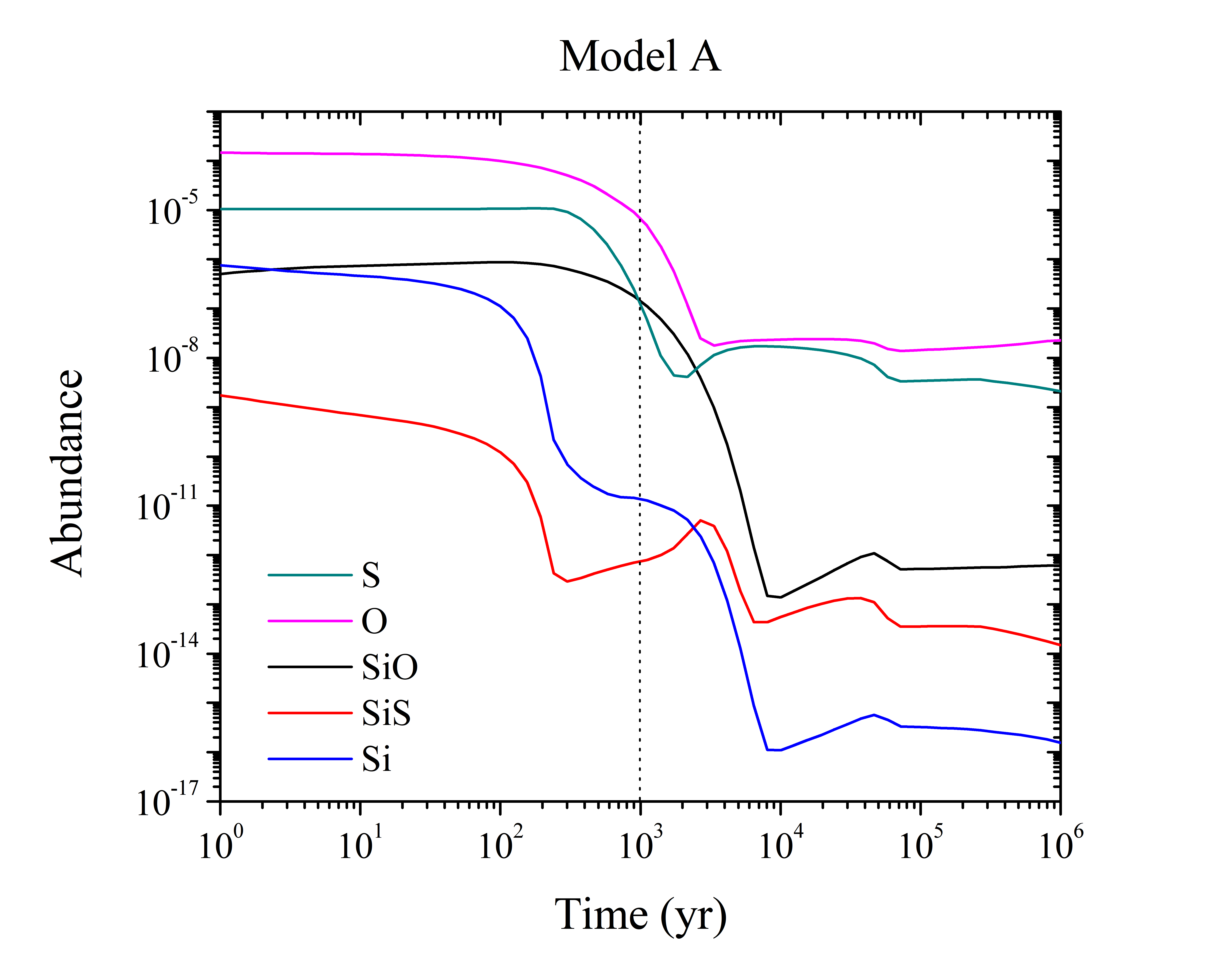}\\
\includegraphics[width=0.47\textwidth,keepaspectratio]{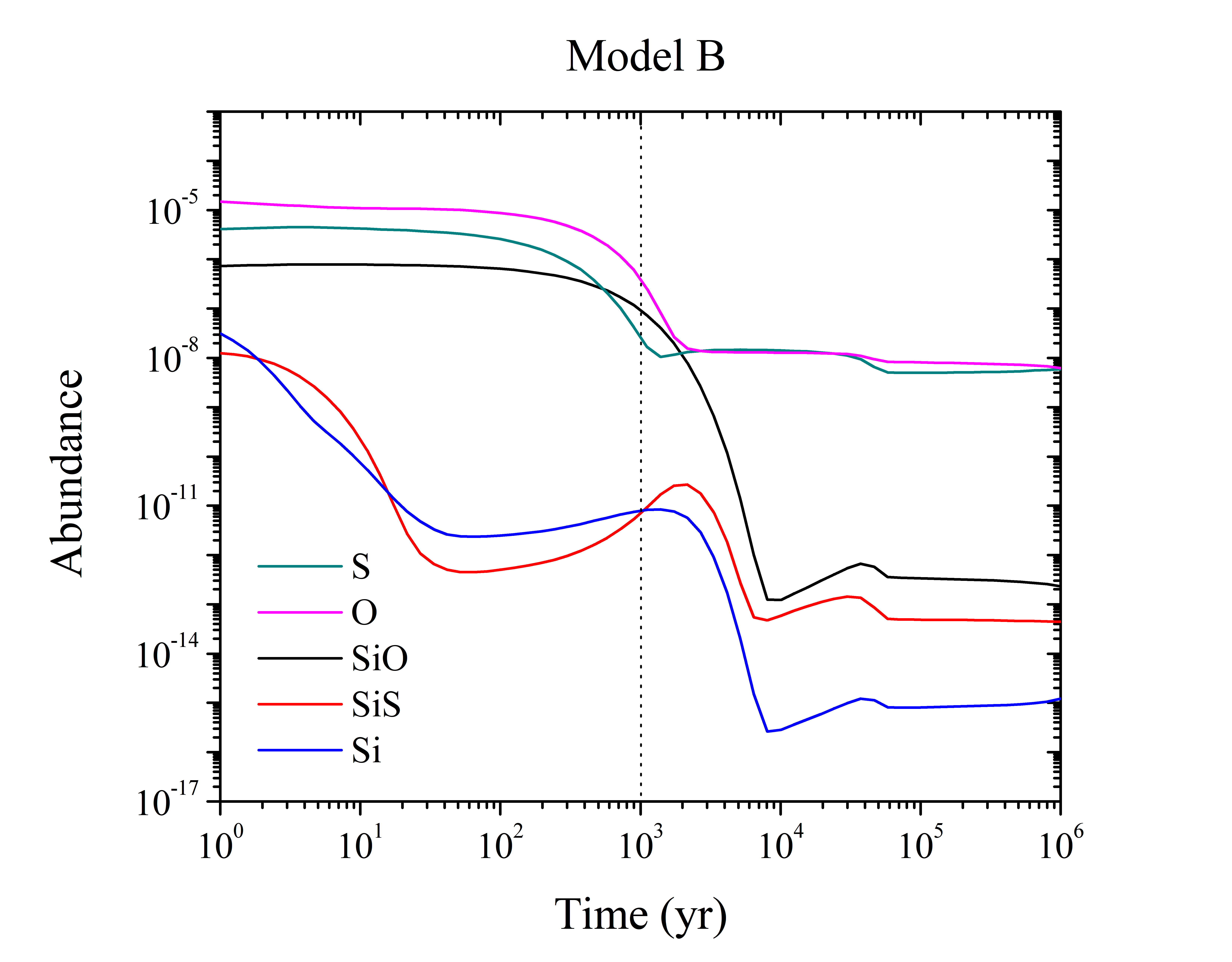}\\
\includegraphics[width=0.47\textwidth,keepaspectratio]{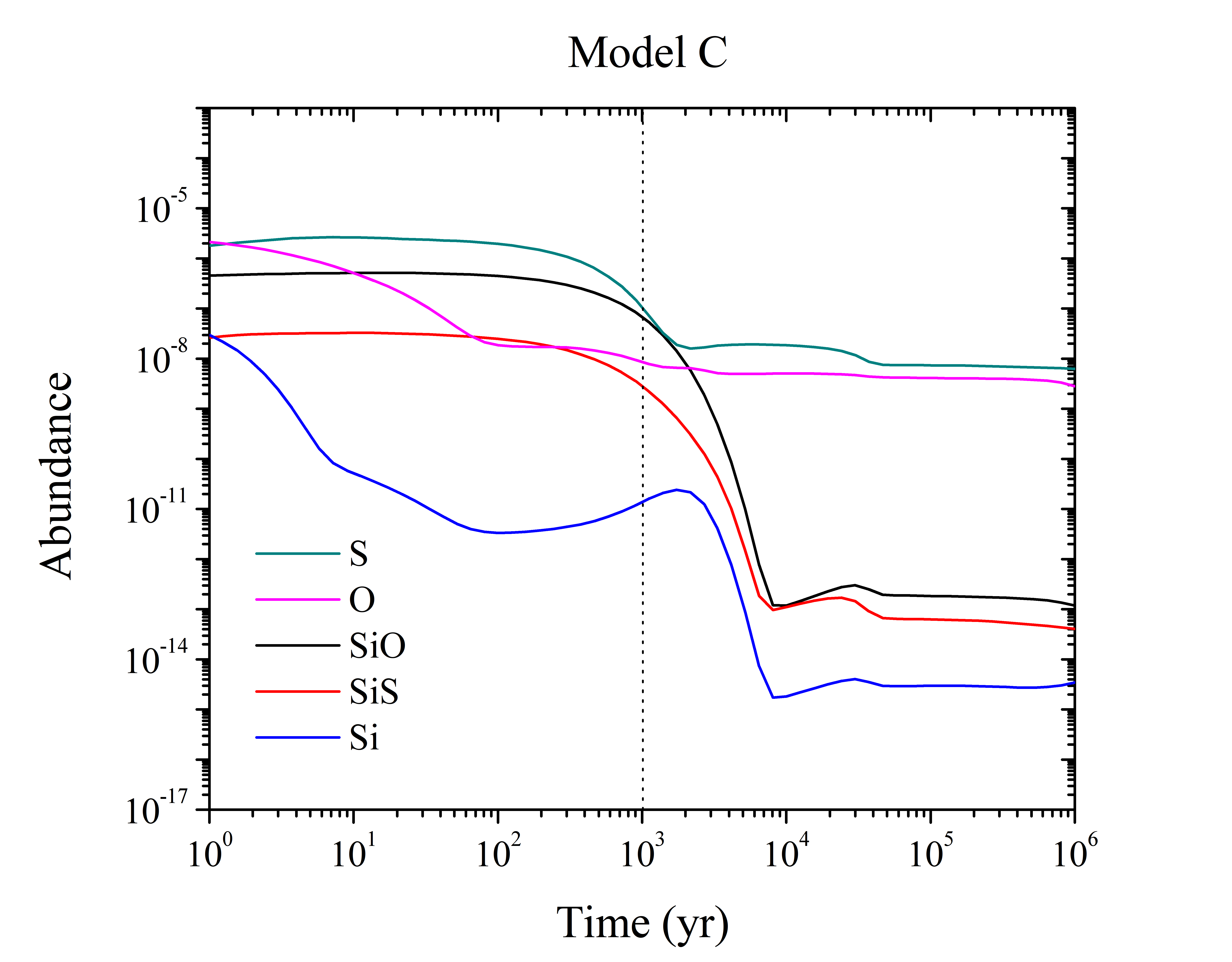}
\caption{\label{fig:Graph5} Abundances as a function of time of the chemical elements S, O, and Si, and the molecules SiS and SiO, in a protostellar shock model evolving from a cloud of a) 1 $\times$ 10$^5$, b) 5 $\times$ 10$^5$ and c) 1 $\times$ 10$^6$ yr. The dotted-vertical lines are present to merely guide the eye regarding the dynamical age of a  shock template source (See the text in \S~\ref{sec:6}).}
\end{center}
\end{figure}

The astrochemical models are presented in Fig.~\ref{fig:Graph5}, where three simulations are exhibited for shocks evolving from clouds of 0.1, 0.5 and 1 $\times$ 10$^6$~yr. In this figure it is shown the abundances of S, O, Si, SiO and SiS since the beginning of the 3rd step up to t=10$^6$~yr, where the stationary equilibrium takes place in the gas phase. Two different effects can be noted from the neutrals SiO and SiS: while the SiO abundance remains high and \ \lq invariable\rq \ in the different models, the SiS one increments systematically with the cloud lifetime. The SiS abundance increases almost three orders of magnitude when the cloud lifetime goes from  1 $\times$ 10$^5$ to 1 $\times$ 10$^6$~yr. In part, the responsible mechanisms for such SiS production are the reactions Si + H$_2$S $\longrightarrow$ SiS + H$_2$, Si + HS $\longrightarrow$ SiS + S, listed in Table~\ref{tab:net}, and the electronic recombination reaction HSiS$^+$ + e$^-$ $\longrightarrow$ H + SiS. Another factor that favours the SiS accumulation is the depletion of O, as evidenced by the pink curve in Fig.~\ref{fig:Graph5}, since it  destroys SiS via SiS + O $\longrightarrow$ SiO +S.

\subsection{Impact of the new chemical reactions}
\label{sec:6.1}

\begin{figure}
\begin{center}
\includegraphics[width=0.47\textwidth,keepaspectratio]{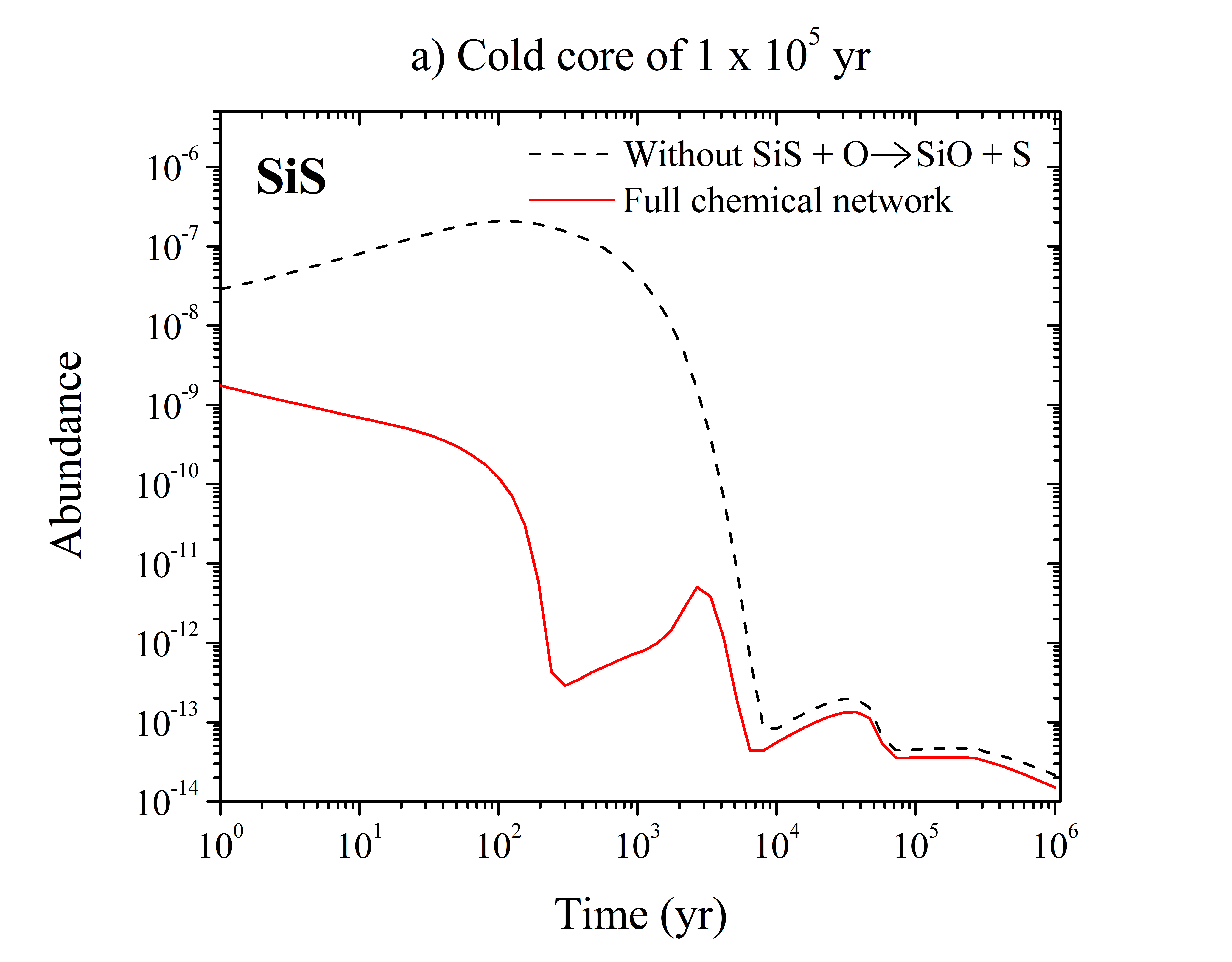}\\
\includegraphics[width=0.47\textwidth,keepaspectratio]{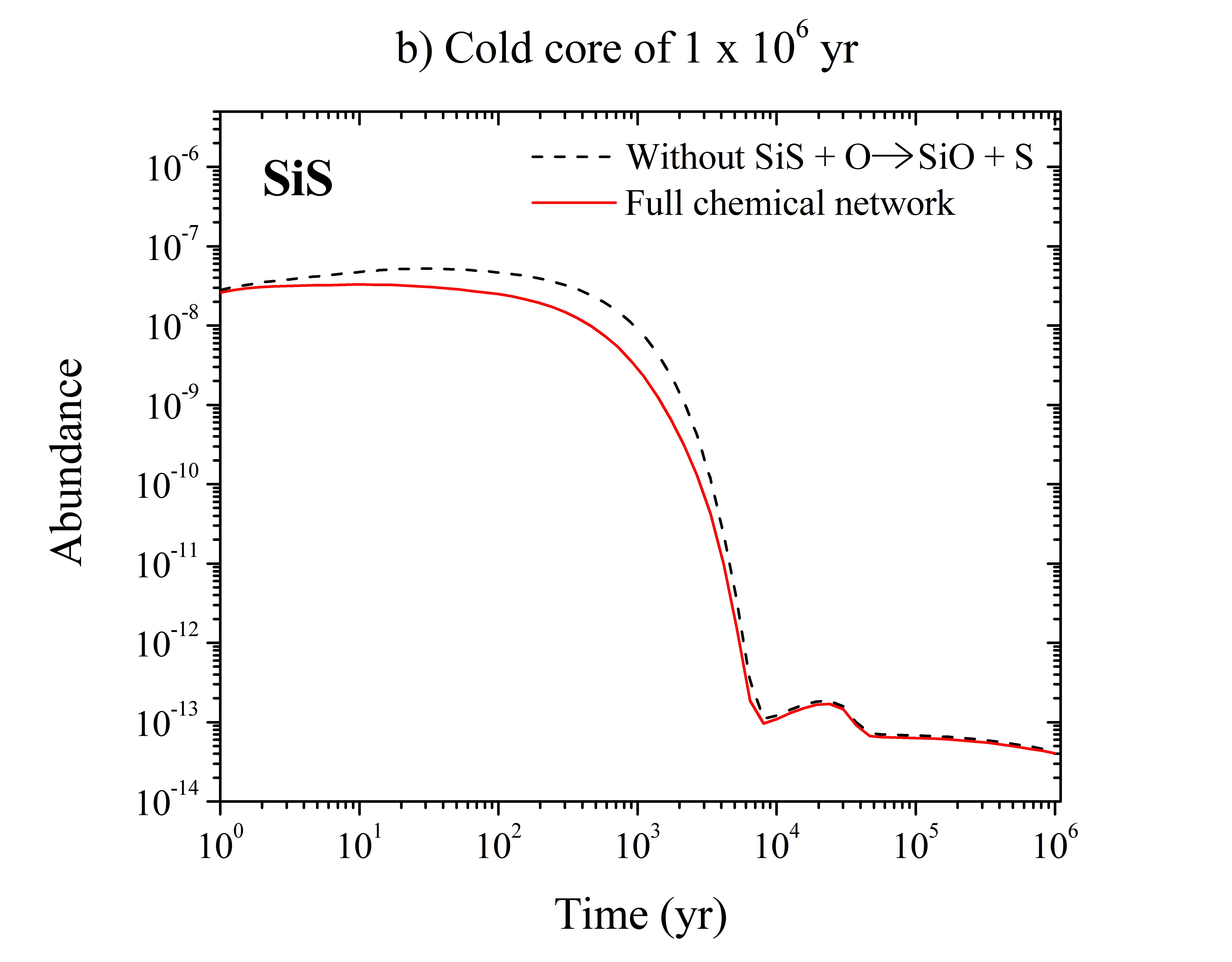}
\caption{\label{fig:Graph6} SiS 
abundance with and without including the destruction reaction SiS + O $\longrightarrow$ SiO + S in a shock model evolving from a cloud of a) 1 $\times$ 10$^5$ and b) 1 $\times$ 10$^6$~yr.}
\end{center}
\end{figure}

Based on the results presented by \citet{MOT21:37}, obtained in the context of a protostellar shock region, our goal here has been to expand such approach by computing a) new chemical reactions (Table~\ref{tab:net}) and b) testing different  cloud lifetimes (\S~\ref{Sec:5.1}).  As a result, the inclusion of the new chemical reactions, in the shock model rising from a cloud of 1 $\times$ 10$^5$~yr, diminished the SiS abundances. Under the same conditions, but allowing a longer cloud evolution of 1 $\times$ 10$^6$~yr, it is observed that the SiS abundance reaches again the observed value, of the order of 10$^{-8}$ \citep{podio2017silicon}. In Fig.~\ref{fig:Graph6}, the destruction effects of SiS by O are evidenced. SiS is easily destroyed in the younger models due to the predominance of the destruction reaction SiS + O $\longrightarrow$ SiO + S, responsible for  diminishing the SiS abundance in the shock model around 10$^3$~yr. In contrast, the SiS abundance is less affected as the cloud lifetime gets older, since this allows a higher accumulation of SiS so that the destruction reaction losses its predominance.

\begin{figure*}
\begin{center}
\includegraphics[width=0.47\textwidth,keepaspectratio]{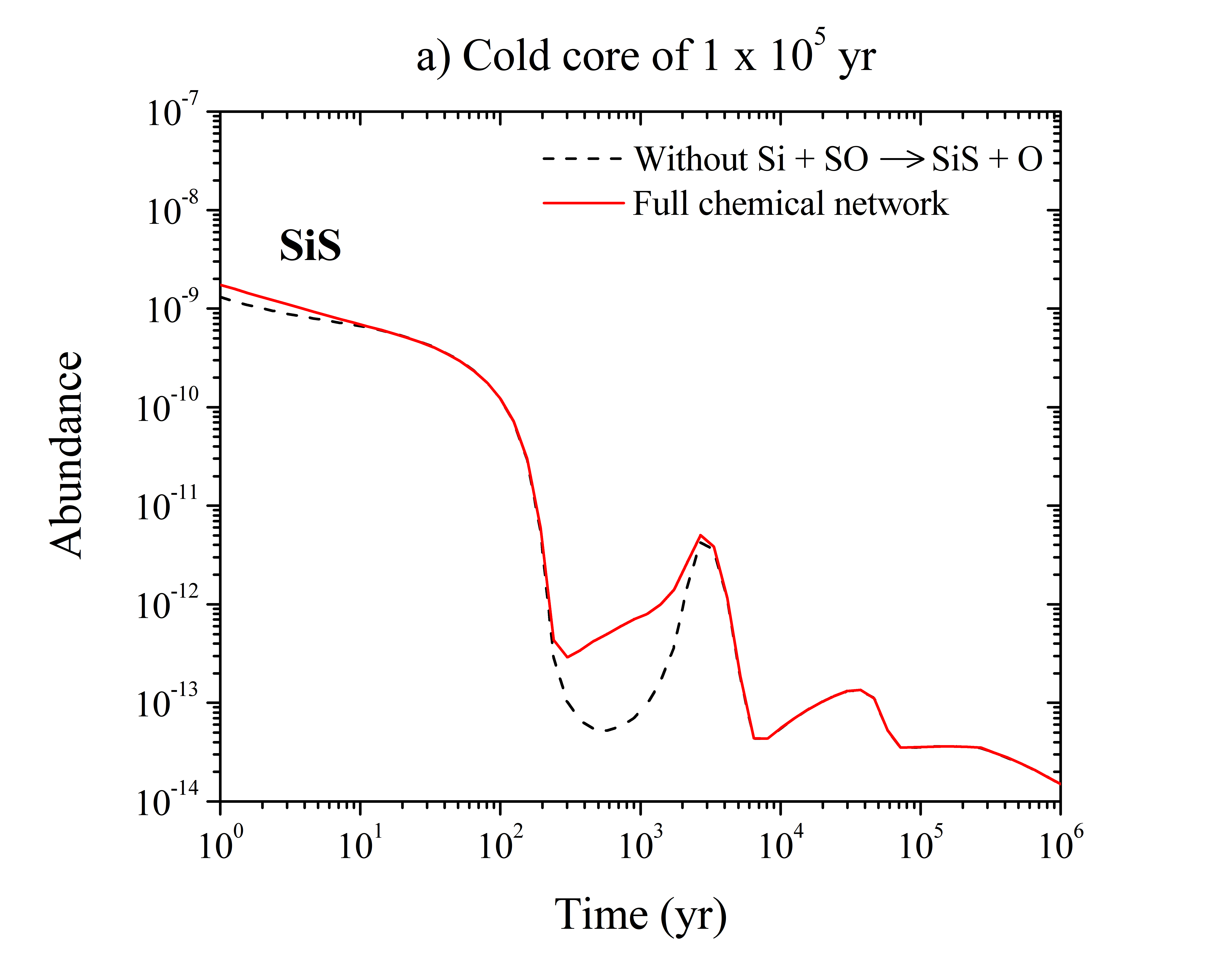}
\includegraphics[width=0.47\textwidth,keepaspectratio]{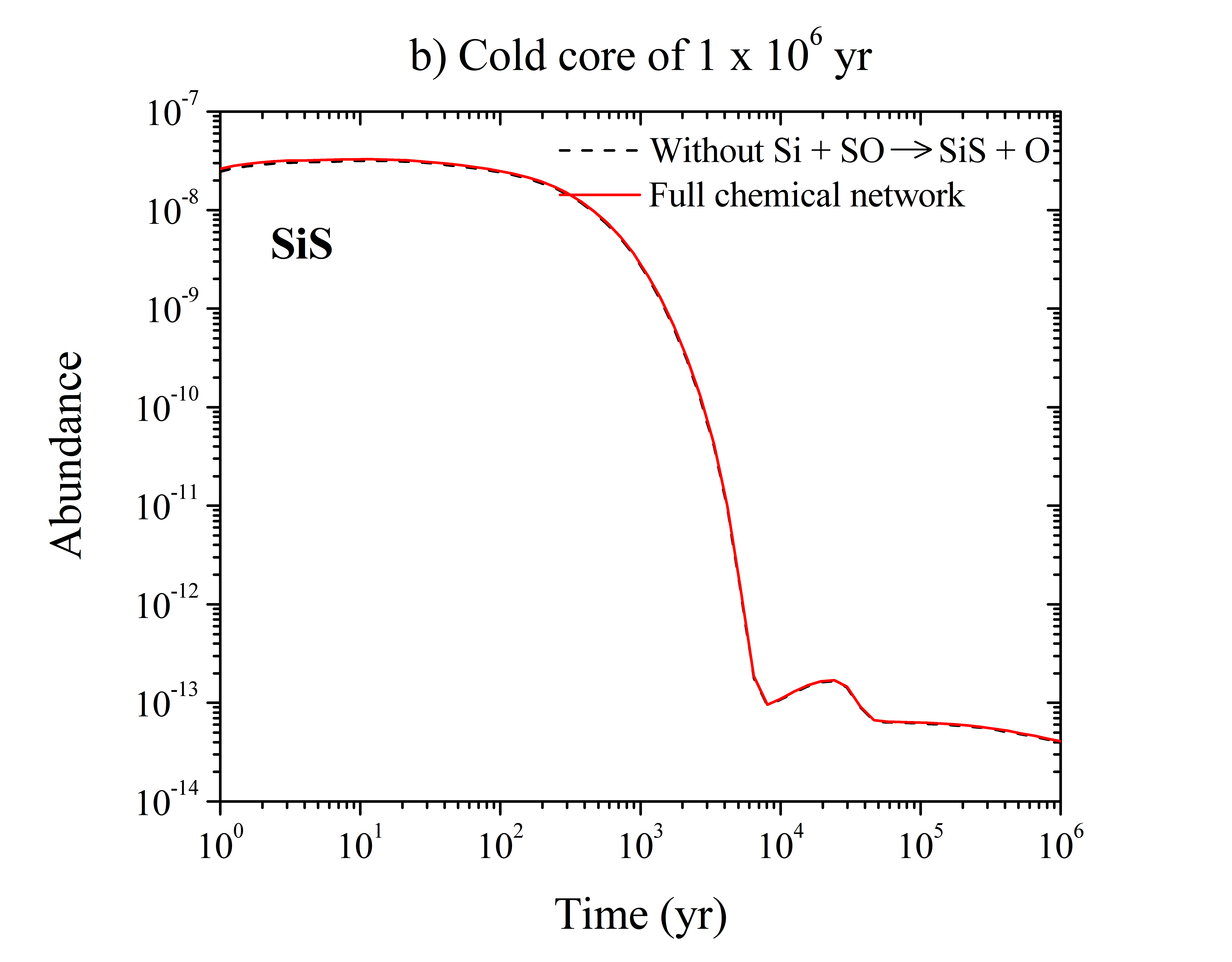}
\includegraphics[width=0.47\textwidth,keepaspectratio]{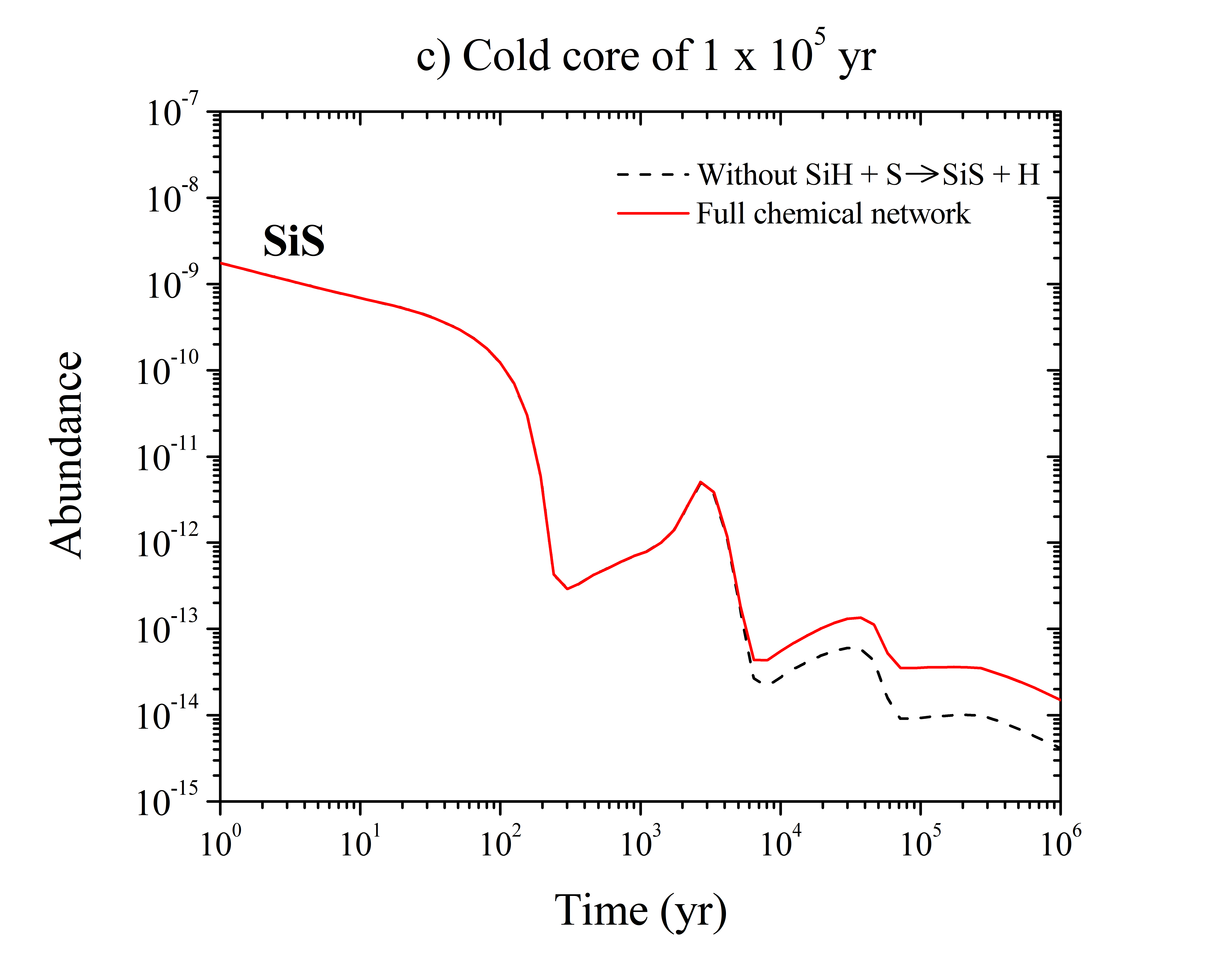}
\includegraphics[width=0.47\textwidth,keepaspectratio]{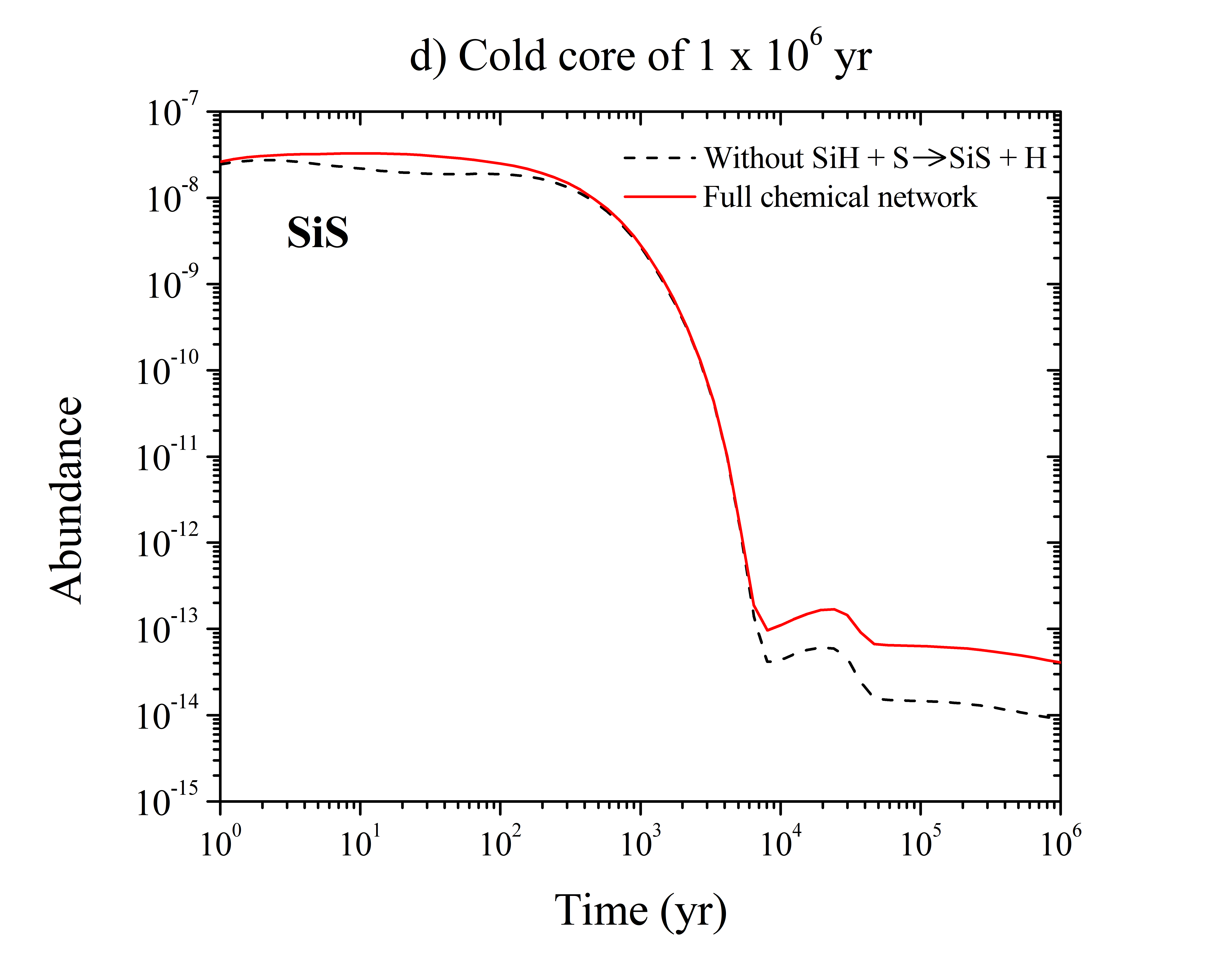}
\caption{\label{fig:Graph7} SiS 
abundance with and without including the production reaction Si + SO $\longrightarrow$ SiS + O in a shock model evolving from a cloud of a) 1 $\times$ 10$^5$ and b) 1 $\times$ 10$^6$~yr. Similarly, panels c) and d) stand for the SiH + S $\longrightarrow$ SiS + H reaction.}
\end{center}
\end{figure*}

In order to evaluate the relevance of other production mechanisms, we show in  Fig.~\ref{fig:Graph7} the effects of excluding the ($i.$) Si + O $\longrightarrow$ SiS + O, and ($ii.$) SiH + S $\longrightarrow$ SiS + H reactions. In both cases, the major effects on the SiS abundance are observed in the youngest model. For which, the absence of $i.$ is noticeable at ages between 2 $\times$ 10$^2$ and 2 $\times$ 10$^3$~yr; in contrast, the absence of $ii.$ is noticeable after 3 $\times$ 10$^3$~yr, which is late if we consider the dynamic and chemical timescale of L1157-B1, of the order of 10$^3$~yr.
There are open questions that warrant further studies on the SiS chemistry: how efficient is the SiS formation in the circumstellar envelopes of evolved stars? Is that  comparable with the case of outflows driven by young stellar objects? Thus, follow-up studies will permit us deeper discussions about the presence of SiS and other Si- and S-bearing molecules in different astronomical objects. 

\subsection{Comparison with observational studies}

SiS has been observed in a few astronomical objects: sources harbouring outflows and circumstellar envelopes, among them. The first SiS detection was reported in the circumstellar envelope of IRC+10216 \citep{Morris1975}; including a maser emission detection in a most recent work \citep{Gong2017}.  \citet{Velilla2017} observed rotational lines both of SiS and the isotopologues $^{29}$SiS, $^{30}$SiS, Si$^{34}$S towards IK Tau, a miratype variable star surrounded by a O-rich circumstellar envelope. The results reported by them suggest that SiO/SiS $\gtrsim$ 2.  In a previous work, \citet{Gobrecht2016} developed a chemical model based on the same source, predicting the abundance values SiS $\sim$ (2.2--2.7) $\times$ 10$^{-7}$ and SiO $\sim$ (1.9--6.4) $\times$ 10$^{-5}$.  \citet{Ziurys1988,Ziurys1991} studied the importance of SiS as a tracer for the chemistry of outflows, finding that  SiO/SiS $\sim$ 40--80 in sources harbouring outflows, such as W51, Orion-S, Orion-KL, Sgr B2(N) and W49. In a survey towards L1157-B1, \citet{POD17:L16} found an abundance gradient given by the SiO/SiS ratio in L1157-B1 and -B2. They reported ratios ranging from $\sim$~25 to 180. In comparison with our models, we found the abundance ratio SiO/SiS $\sim$ 23, at an age of 1000~yr, obtained from the shock model that evolves from a cloud of 10$^6$~yr. That result represents our best approximation of the  calculated SiS abundance under the physical and chemical conditions of a protostellar shocked gas. Despite the predicted and observed ratios are similar, an important aspect is to scrutinise how the cloud lifetime affects the molecular abundances in shocks, for instance, testing timescales of 10$^5$--10$^6$~yr. Similar implications were found in observations and models of HNCO, another tracer associated with outflow dynamics and chemistry (e.g. \citealt{Rodriguez2010,Velilla2015,Canelo2021}).

\section{Conclusions and perspectives}
\label{sec:conc}


In this work we have performed an accurate computational study of the reactions involving the Si + SO$_2$ chemical system. These calculations, coupled with other data available in the literature, allowed us to achieve several important results listed below:

\begin{enumerate}
    \item The \ce{Si +SO2} collision will quickly produce  \ce{SiO +SO}, and no other products are kinetically available in the literature.
    \item  In \ce{S +SiO2} and \ce{O +OSiS} collisions, \ce{SiO +SO} will also be quickly formed. 
    \item SiS is kinetically stable with respect to collisions with \ce{O2}.
    \item The chemical reactions studied here were tested in a simplified model of a protostellar shock using the {\sc Nautilus} gas-grain code. The model consisted in simulating the physical and chemical conditions of a shocked gas that evolves from its primeval cold core, then the shock region by itself, and finally to mimic the gas bulk conditions after the passage of the shock. 
    \item The results indicate that neutral-neutral reactions, previously not included in the major astrochemical databases, have a profound impact in the study of the abundances of S- and Si-bearing molecules. A particular emphasis to the destruction of SiS by collisions with atomic oxygen has been carried out. In addition, the lifetime of the primeval molecular cloud prior to the shock have a relevant role in the chemistry; models assuming the oldest cloud ages, of the order of 10$^6$~yr, provided a better match between the observed and modelled abundances of SiS, of SiS/H$_2$ $\thickapprox 10^{-8}$ in a protostellar shock as L1157-B1. We took that source as an example because SiS was recently detected towards it.
    \item The \ce{SiH +S} reaction have a significant role  on SiS production in the later stages after the passage of the shock.
    \item Molecules such as SiS have been detected in a few astronomical sources, i.e., protostellar objects harbouring molecular outflows and circumstellar envelopes of evolved AGB stars. Follow-up studies combining observational and theoretical approaches will be carried out to explore more deeply the presence and abundance of Si- and S-bearing molecules in regions of the ISM.
\end{enumerate}

\section*{Acknowledgements}

The authors would like to thank the financial support provided  by the Coordena\c c\~ao de Aperfei\c coamento de Pessoal de N\'ivel Superior - Brasil (CAPES) - Finance Code 001,
Conselho Nacional de Desenvolvimento Cient\'ifico e Tecnol\'ogico (CNPq), grant 311508/2021-9, and
Funda\c c\~ao de Amparo \`a Pesquisa do estado de Minas Gerais (FAPEMIG).
Rede Mineira de Química (RQ-MG) and CEFET-MG are also acknowledged. We are also thankful for computational resources provided by LNCC and the Santos Dumont supercomputer.
EM acknowledges support under the grant "Maria Zambrano" from the UHU funded by the Spanish Ministry of Universities and the "European Union NextGenerationEU". This project has also received funding from the European Union's Horizon  2020 research and innovation program under the Marie  Sk{\l}odowska-Curie grant agreement No 872081 and from grant  PID2019-104002GB-C21 funded by MCIN/AEI/ 10.13039/501100011033 and,  as appropriate, by “ERDF A way of making Europe”, by the “European  Union” or by the “European Union NextGenerationEU/PRTR”.  This work  has also been partially supported by the Consejer\'{\i}a de Transformaci\'on Econ\'omica, Industria, Conocimiento y Universidades, Junta de Andaluc\'{\i}a and European Regional Development Fund (ERDF 2014-2020) PY2000764, and by the Ministerio de Ciencia, Innovaci\'on y  Universidades (ref.COOPB20364).

\section*{DATA AVAILABILITY}
The data underlying this article are available in the article and in its online supplementary material.




\clearpage

\bibliographystyle{mnras}
\bibliography{SiS}






\bsp	
\label{lastpage}
\end{document}